\documentclass{ws-procs9x6}            % default, citations in superscript
\begin{document}
\title{Canonical interpretation of the $X(4140)$ state within the $^3P_0$ model}

\author{Wei Hao, Guan-Ying Wang, En Wang$^*$, De-Min Li and Guan-Nan Li}
\address{School of Physics and Microelectronics, Zhengzhou University, Zhengzhou, Henan 450001, China\\
$^*$E-mail: wangen@zzu.edu.cn}

\begin{abstract}
Recently, LHCb collaboration has confirmed the state $X(4140)$, with a much larger width $\Gamma= 83\pm 21^{+21}_{-14}$~MeV than the previous experimental measurements, which has confused the understanding of its nature. We have investigated the possibility of the $\chi_{c1}(3P)$ interpretation for the $X(4140)$, considering the mass and the strong decay properties.
\end{abstract}


\bodymatter

\section{Introduction}
Recently, the  LHCb measurements has confirmed the $X(4140)$ state with high statistic data~\cite{Aaij:2016iza,Aaij:2016nsc},  with a mass $4146.5\pm 4.5^{+4.6}_{-2.8}$~MeV and a width $ 83\pm 21^{+21}_{-14}$~MeV, much larger than the previous experimental measurements~\cite{PDG2018}, and the quantum numbers were determined to be $J^{PC}=1^{++}$.
There are many different suggestions about the structure of the $X(4140)$~\cite{Chen:2016iua,Wang:2018qpe}, especially because of the large discrepancy of the width.

Indeed, it is natural and necessary to exhaust the possible $q\bar{q}$ description of the observed states before restoring to the more exotic assignments.
 In this work, taking the meson wave functions by solving the relativistic/non-relativistic Schr\"{o}dinger equation, we investigate the decay properties of the $X(4140)$ as the assignment of charmonium state in the $^3P_0$ model, and provide more information about the decay modes for searching for the $X(4140)$, which could be useful to extract the more precision width.

\section{The $^3P_0$ decay model}
\label{sec:model}
The $^3P_0$ model, also known as the quark-pair creation model, has been widely applied to study strong decays of hadrons with considerable success~\cite{Micu:1968mk,Close:2005se,Li:2009rka,Lu:2016bbk}.
In the model, the transition operator $T$ of the decay $A\rightarrow BC$ in the $^3P_0$ model can be written,
\begin{eqnarray}
  T &=&-3\gamma\sum_m\langle 1m1-m|00\rangle\int
    d^3\boldsymbol{p}_3d^3\boldsymbol{p}_4\delta^3(\boldsymbol{p}_3+\boldsymbol{p}_4)\nonumber\\
&& \times   {\cal{Y}}^m_1\left(\frac{\boldsymbol{p}_3-\boldsymbol{p}_4}{2}\right
    )\chi^{34}_{1,-m}\phi^{34}_0\omega^{34}_0b^\dagger_3(\boldsymbol{p}_3)d^\dagger_4(\boldsymbol{p}_4),
\end{eqnarray}
where $\gamma$ is a dimensionless parameter corresponding the strength of quark-antiquark $q_3\bar{q}_4$ pair producing from the vacuum.
The decay width
$\Gamma(A\rightarrow BC)$ can be expressed in terms of the partial wave amplitude,
\begin{eqnarray}
\Gamma(A\rightarrow BC)= \frac{\pi
|\boldsymbol{P}|}{4M^2_A}\sum_{LS}|{\cal{M}}^{LS}(\boldsymbol{P})|^2, \label{width1}
\end{eqnarray}
where the explicit expressions for ${\cal{M}}^{LS}(\boldsymbol{P})$ can be found in Refs.~\citenum{Close:2005se,Li:2009rka,Lu:2016bbk}.

In this work, we  take into account two choices of the wave functions for mesons: Case A: the modified non-relativistic quark model (MNRQM), Case B: the modified  Godfrey-Isgur model (MGI). The details of model and the wave functions can be found in Ref.~\citenum{Hao:2019fjg}.

\section{Results and discussions}
\label{sec:results}

The mass spectra of the charmonium states predicted by two kinds of quark models are shown in Ref.~\citenum{Hao:2019fjg}. According to the PDG~\cite{PDG2018}, the mass of the $X(4140)$ [$I^G(J^{PC})=0^+(1^{++})$] is $4146.8\pm 2.4$~MeV, which is consistent with the predicted mass of $\chi_{c1}(3P)$ in both models within the uncertainties of the models. Next, we will calculate the strong decay widths of the $X(4140)$ state as the $\chi_{c1}(3P)$  assignment.

In the $^3P_0$ model, the strength of quark-antiquark pair creating from the vacuum, is taken to be $\gamma=4.52$ in Case A,  and $\gamma=5.90$ for Case B, by fitting to the total widths of the well established charmonium states,  $\psi(3770)$  ($1^3D_1$), $\psi(4040)$ ($3^3S_1$), $\psi(4160)$ ($2^3D_1$), and $\chi_{c2}(2P)$.
\begin{table}%[htpb]
\begin{center}\label{tab:width}
\tbl{Decay widths of the ${\chi}_{c0}(3P)$, ${\chi}_{c1}(3P)$ and ${\chi}_{c2}(3P)$ states (in MeV). The mass of the ${\chi}_{c1}(3P)$ is taken to be the one of the $X(4140)$, and the masses of the ${\chi}_{c0}(3P)$,  and ${\chi}_{c2}(3P)$ are taken from Ref.~\citenum{Hao:2019fjg}, respectively for Case A and Case B.}
%\footnotesize
{\begin{tabular}{c|c|c|c|c}
\hline\hline
 State                &Channel                        &Mode                 &$\Gamma$ (Case A)  &$\Gamma$ (Case B)\\
\hline
${\chi}_{c0}(3P)$     &$0^+\rightarrow 0^-0^-$           &$D\bar{D}$           &10.58          &0.22    \\
    $$                &$$                             &$D^+_sD^-_s$           &0.37            &1.87    \\
    $$                &$0^+\rightarrow 1^-1^-$            &$D^*\bar{D}^*$       &16.28           &35.95  \\
Total Width           &$$                             &$$                    &27.23          &38.03  \\
\hline
${\chi}_{c1}(3P)$     &$1^+\rightarrow 0^-1^-$          &$D\bar{D}^*$         &4.54           &14.48     \\
    $$                &$$                             &$D_s\bar{D}_s^*$     &1.23            &0.70    \\
    $$                &$1^+\rightarrow 1^-1^-$           &$D^*\bar{D}^*$       &6.86           &16.17   \\
Total Width           &$$                             &$$                    &12.63          &31.34    \\
\hline
${\chi}_{c2}(3P)$     &$2^+\rightarrow 0^-0^-$           &$D\bar{D}$           &7.71            &8.79      \\
    $$                &$$                             &$D^+_sD^-_s$           &0.63            &0.10 \\
    $$                &$2^+\rightarrow 0^-1^-$           &$D\bar{D}^*$         &20.04            &11.34      \\
    $$                &$$                             &$D_s\bar{D}_s^*$     &0.17            &0.13     \\
    $$                &$2^+\rightarrow 1^-1^-$           &$D^*\bar{D}^*$       &11.33            &26.87    \\
Total Width          &$$                             &$$                    &39.89         &47.23    \\
\hline\hline
\end{tabular}}
\end{center}
\end{table}

\begin{figure}[htpb]
  \centering
  \includegraphics[scale=0.33]{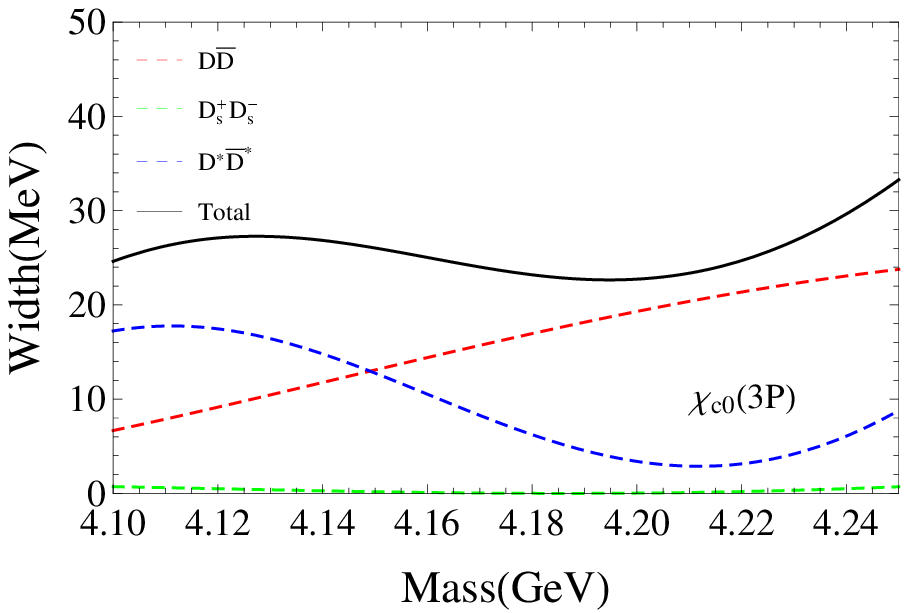}
  \includegraphics[scale=0.33]{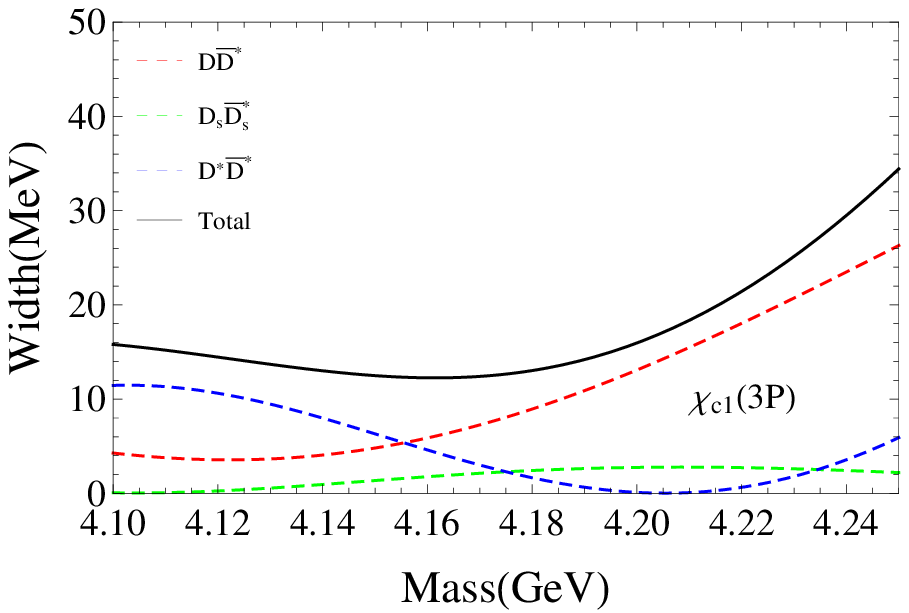}
  \includegraphics[scale=0.33]{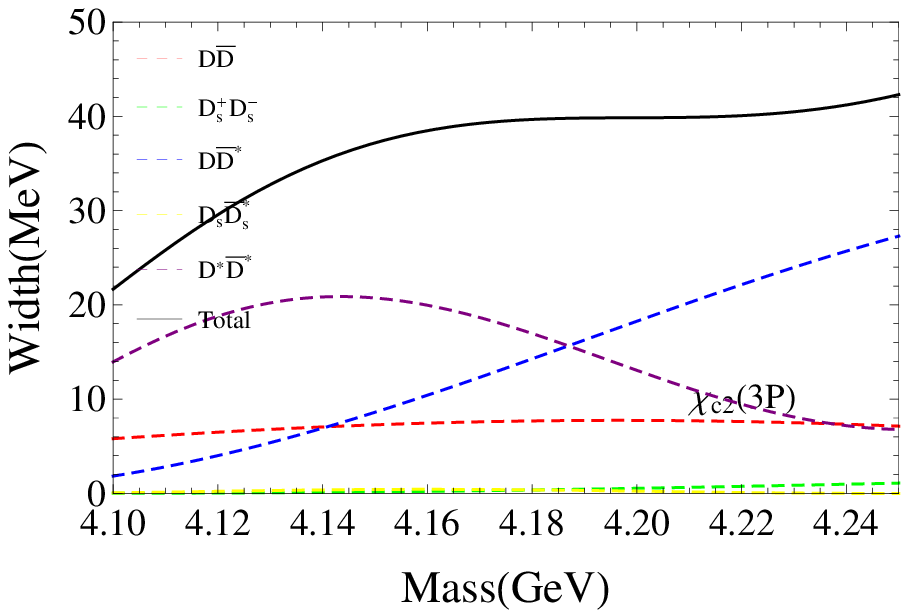}\\
  \includegraphics[scale=0.33]{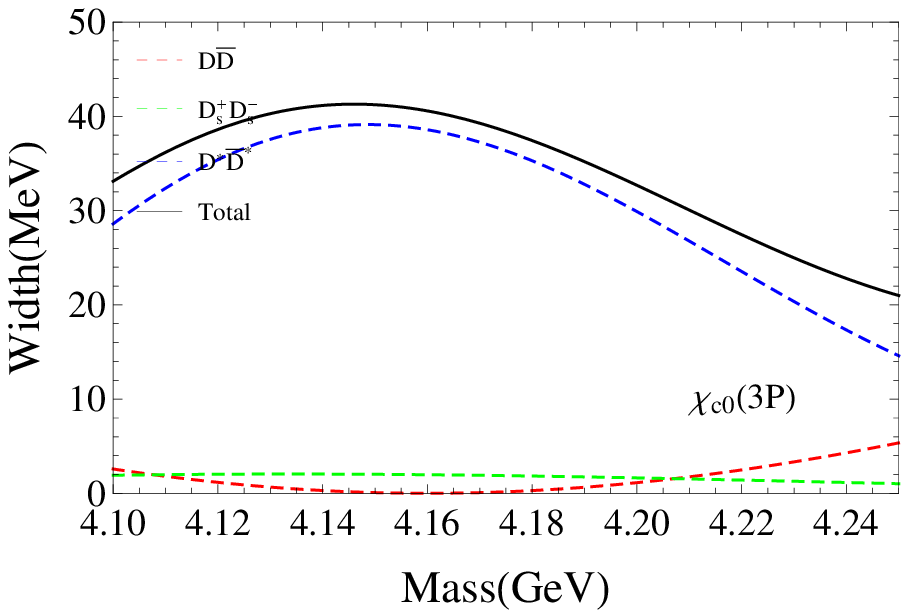}
\includegraphics[scale=0.33]{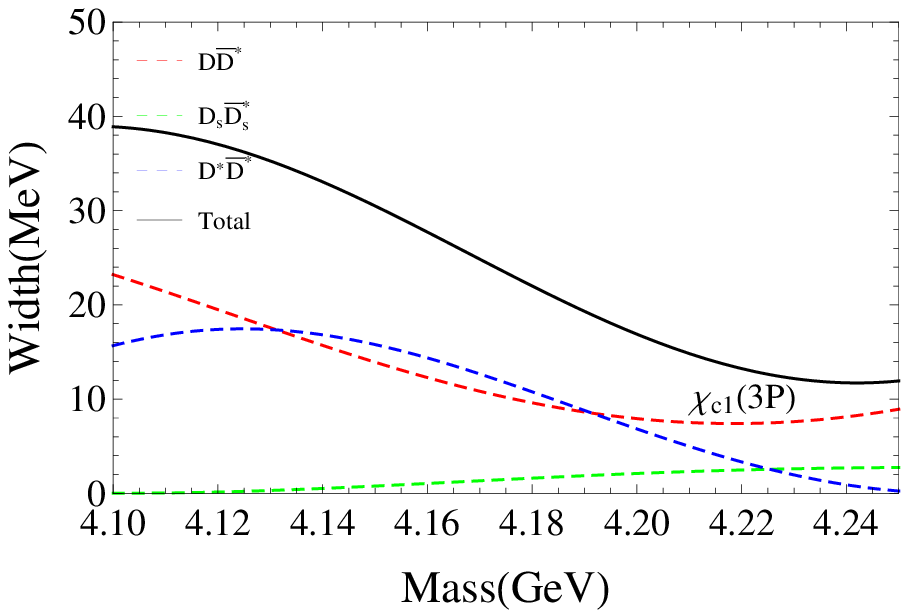}
\includegraphics[scale=0.33]{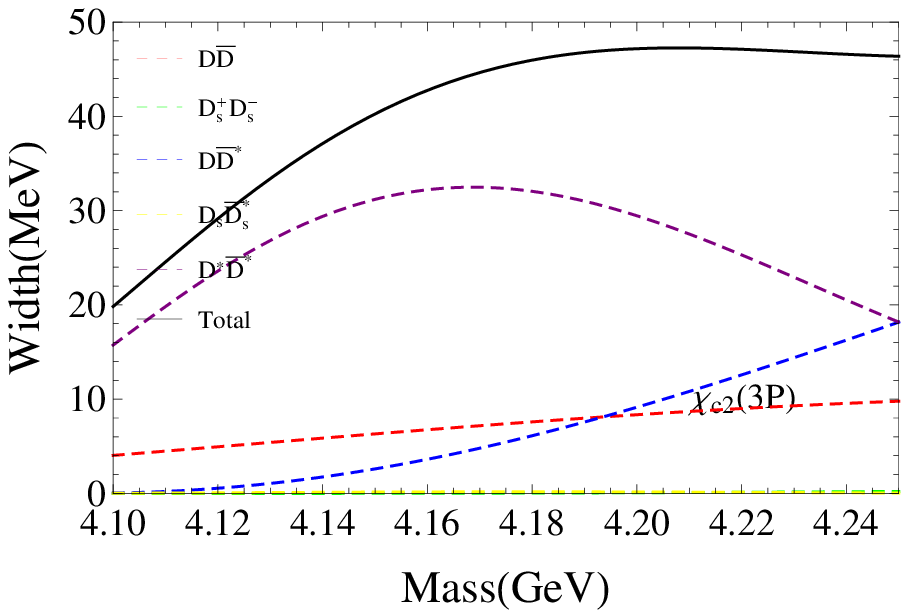}
  \caption{The dependence of the width of ${\chi}_{c0}(3P)$, ${\chi}_{c1}(3P)$ and ${\chi}_{c2}(3P)$ on the initial state mass with the wave functions of Case A (up panels) and Case B (low panels).}
  \label{fig:widthNRQM}
\end{figure}

With the above parameters, we have calculated the partial decay widths and total decay width, as shown in Table~\ref{tab:width} for both Case A and Case B. The total widths are 12.63~MeV for Case A, and $31.34$~MeV for Case B, which are consistent with the average value $\Gamma=22^{+8}_{-7}$~MeV of PDG, considering the uncertainties of the $^3P_0$ model.  It should be pointed out that the decay modes $D\bar{D}^*$ and $D^*\bar{D}^*$ have large decay widths. We suggest to search for this state in those two channels, and to measure the width precisely, which can be shed light on its nature.

We also show the dependence of the $\chi_{c1}(3P)$ decay width on the masses in Fig.~\ref{fig:widthNRQM}. Indeed, if the small width of the $X(4140)$ is confirmed in future high-precision measurements, the $X(4140)$ could be explained as the charmonium state $\chi_{c1}(3P)$. In addition, the processes of the $B^+\to  J/\psi \phi K^+$  and $e^+e^- \to \gamma J/\psi \phi K^+$ decay were investigated in Refs.~\citenum{Wang:2017mrt,Wang:2018djr} ,
where the existences of the $X(4140)$, with the small width, and the molecular state $X(4160)$ is supported by the low $J/\psi \phi$  invariant mass distributions ~\cite{Aaij:2016iza,Aaij:2016nsc,Ablikim:2017cbv}.

Finally, the strong decay width of the $\chi_{c0}(3P)$ and $\chi_{c2}(3P)$ states in Table~\ref{tab:width}, and the mass dependence of the total widths are also shown in Fig.~\ref{fig:widthNRQM}.

\section{SUMMARY}
\label{sec:summary}
We have investigated the strong decay properties of the $X(4140)$ with the assignment of the $\chi_{c1}(3P)$ states in the $^3P_0$ model, taking two cases for the wave functions.
The total decay width of the $\chi_{c1}(3P)$ is predicted to be $12.63$~MeV for Case A, and $31.34$~MeV for Case B, both of which are in agreement with the PDG average width of $X(4140)$. Thus, we conclude that, the $X(4140)$, with a small width, could be explained as the charmonium state $\chi_{c1}(3P)$, and the high-precision measurement about the $X(4140)$ could shed light on the nature of the $X(4140)$.

\section{Acknowledgements}

This work is partly supported by the National Natural Science Foundation of China under Grant Nos. 11505158, 11605158, the Key Research Projects of Henan Higher Education Institutions (No. 20A140027), and the Academic Improvement Project of Zhengzhou University.

\bibliographystyle{ws-procs9x6} % for numbered citation & references
\bibliography{ws-pro-sample}

\begin{thebibliography}{10}

%\cite{Aaij:2016nsc}
\bibitem{Aaij:2016nsc}
  R.~Aaij {\it et al.} [LHCb Collaboration],
  %Amplitude analysis of $B^+\to J/\psi \phi K^+$ decays,
  Phys.\ Rev.\ D {\bf 95}, 012002 (2017).
%  doi:10.1103/PhysRevD.95.012002
%  [arXiv:1606.07898 [hep-ex]].
  %%CITATION = doi:10.1103/PhysRevD.95.012002;%%
  %50 citations counted in INSPIRE as of 18 Sep 2017

%\cite{Aaij:2016iza}
\bibitem{Aaij:2016iza}
  R.~Aaij {\it et al.} [LHCb Collaboration],
  %Observation of $J/\psi\phi$ structures consistent with exotic states from amplitude analysis of $B^+\to J/\psi \phi K^+$ decays,
  Phys.\ Rev.\ Lett.\  {\bf 118}, 022003 (2017).
%  doi:10.1103/PhysRevLett.118.022003
%  [arXiv:1606.07895 [hep-ex]].
  %%CITATION = doi:10.1103/PhysRevLett.118.022003;%%
  %59 citations counted in INSPIRE as of 18 Sep 2017

%\cite{PDG2018}
\bibitem{PDG2018}
  M.~Tanabashi {\it et al.} [Particle Data Group],
  Review of Particle Physics,
  Phys.\ Rev.\ D {\bf 98}, no. 3, 030001 (2018).
%  doi:10.1103/PhysRevD.98.030001
  %%CITATION = doi:10.1103/PhysRevD.98.030001;%%
  %844 citations counted in INSPIRE as of 17 Jan 2019

%\cite{Chen:2016iua}
\bibitem{Chen:2016iua}
  D.~Y.~Chen,
 % Where are $\chi _{cJ}(3P)$ ?,
  Eur.\ Phys.\ J.\ C {\bf 76}, no. 12, 671 (2016).
%  doi:10.1140/epjc/s10052-016-4531-9
%  [arXiv:1611.00109 [hep-ph]].
  %%CITATION = doi:10.1140/epjc/s10052-016-4531-9;%%
  %11 citations counted in INSPIRE as of 23 Jul 2019

%\cite{Wang:2018qpe}
\bibitem{Wang:2018qpe}
  Z.~G.~Wang and Z.~Y.~Di,
%  Analysis of the mass and width of the $X(4140)$ as axialvector tetraquark state,
  Eur.\ Phys.\ J.\ C {\bf 79}, no. 1, 72 (2019).
%  doi:10.1140/epjc/s10052-019-6596-8
%  [arXiv:1811.12821 [hep-ph]].
  %%CITATION = doi:10.1140/epjc/s10052-019-6596-8;%%
  %5 citations counted in INSPIRE as of 23 Jul 2019
  
  
%\cite{Micu:1968mk}
\bibitem{Micu:1968mk}
  L.~Micu,
 % Decay rates of meson resonances in a quark model,
  Nucl.\ Phys.\ B {\bf 10}, 521 (1969).
  %doi:10.1016/0550-3213(69)90039-X
  %%CITATION = doi:10.1016/0550-3213(69)90039-X;%%
  %317 citations counted in INSPIRE as of 24 Mar 2016
 
%\cite{Close:2005se}
\bibitem{Close:2005se}
    F.~E.~Close and E.~S.~Swanson,
%Dynamics and decay of heavy-light hadrons,
Phys.\ Rev.\ D {\bf 72}, 094004 (2005).

  
%\cite{Li:2009rka}
\bibitem{Li:2009rka}
  D.~M.~Li and E.~Wang,
 % Canonical interpretation of the $\eta_2(1870)$,
  Eur.\ Phys.\ J.\ C {\bf 63},297 (2009).
%  doi:10.1140/epjc/s10052-009-1106-z
 % [arXiv:0904.1252 [hep-ph]].
  %%CITATION = doi:10.1140/epjc/s10052-009-1106-z;%%
  
%\cite{Lu:2016bbk}
 \bibitem{Lu:2016bbk}
   Q.~F.~L\"{u}, T.~T.~Pan, Y.~Y.~Wang, E.~Wang, and D.~M.~Li,
 %Excited bottom and bottom-strange mesons in the quark model,
Phys.\ Rev.\ D {\bf 94},074012 (2016).
      %??doi:10.1103/PhysRevD.94.074012 ??[arXiv:1607.02812 [hep-ph]]. ??%%CITATION = doi:10.1103/PhysRevD.94.074012;%%   %9 citations counted in INSPIRE as of 22 Dec 2017
    
%\cite{Hao:2019fjg}
\bibitem{Hao:2019fjg}
  W.~Hao, G.~Y.~Wang, E.~Wang, G.~N.~Li and D.~M.~Li,
  %``Canonical interpretation of the $X(4140)$ state within the $^3P_0$ model,''
  arXiv:1909.13099 [hep-ph].
  %%CITATION = ARXIV:1909.13099;%%  
    
  
%\cite{Wang:2017mrt}
\bibitem{Wang:2017mrt}
  E.~Wang, J.~J.~Xie, L.~S.~Geng and E.~Oset,
%  Analysis of the $B^+\to J/\psi \phi K^+$ data at low $J/\psi \phi$ invariant masses and the $X(4140)$ and $X(4160)$ resonances,
  Phys.\ Rev.\ D {\bf 97}, no. 1, 014017 (2018).
%  doi:10.1103/PhysRevD.97.014017
%  [arXiv:1710.02061 [hep-ph]].
  %%CITATION = doi:10.1103/PhysRevD.97.014017;%%
  %12 citations counted in INSPIRE as of 12 Aug 2019

%\cite{Wang:2018djr}
\bibitem{Wang:2018djr}
  E.~Wang, J.~J.~Xie, L.~S.~Geng and E.~Oset,
  %``The $X(4140)$ and $X(4160)$ resonances in the $e^+e^-\to \gamma J/\psi \phi $ reaction,''
  Chin.\ Phys.\ C {\bf 43}, no. 11, 113101 (2019).
 % %doi:10.1088/1674-1137/43/11/113101
%  [arXiv:1806.05113 [hep-ph]].
  %%CITATION = doi:10.1088/1674-1137/43/11/113101;%%
  %4 citations counted in INSPIRE as of 19 Nov 2019    

%\cite{Ablikim:2017cbv}
\bibitem{Ablikim:2017cbv}
  M.~Ablikim {\it et al.} [BESIII Collaboration],
  %Observation of $e^{+}e^{-} \to \phi\chi_{c1}$ and $\phi\chi_{c2}$ at $\sqrt{s}$=4.600 GeV,
  Phys.\ Rev.\ D {\bf 97}, no. 3, 032008 (2018).
%  doi:10.1103/PhysRevD.97.032008
%  [arXiv:1712.09240 [hep-ex]].
  %%CITATION = doi:10.1103/PhysRevD.97.032008;%%
  %8 citations counted in INSPIRE as of 13 Nov 2019


\end{thebibliography}

%\end{document}

%Non BiBTeX users can list down their references as:

\end{document}